# The cosmic origin of supersymmetry and internal symmetry

Ding-Yu Chung

In this paper, the cosmic vacuum is the wavefunction of the eleven dimensional Planck supermembrane.  The Planck wavefunction is the superposition of dimensions from eleven to four dimensional spacetime with decreasing energy and increasing size.  The cosmic vacuum is a gigantic cosmic particle-wave.  It undergoes a gigantic slow cosmic oscillation between the high-energy eleven dimensional spacetime and the low-energy four dimensional spacetime.  The origin of the ordinary (baryonic) matter in the big bang universe is the collapsed Planck wavefunction due to the lepton-quark entangled state with two unequal sets of spacetime as the quantum system-measurement entangled state in Hilbert space.  The collapsed Planck wavefunction has four dimensional spacetime and seven dimensional internal space (non-spacetime) with seven gauge bosons. The remnants of the cosmic oscillation are cosmic radiation, gravity, and microscopic particle-wave in the four dimensional spacetime.  Internal symmetries for gauge bosons, leptons, quarks, and black holes are in the seven dimensional internal space.  The result is the big bang universe with the cosmic vacuum.  There is no curled-up space dimension and compacitification. Except at the very beginning of the cosmic cycle and black hole, the big bang universe has pure four-dimensional spacetime. The cosmic vacuum has the four-to-eleven dimensional spacetime supersymmetry, while the big bang universe has the four dimensional spacetime and the seven dimensional internal space. The masses of elementary particles can be calculated with only four known constants: the number of the extra spatial dimensions in the supermembrane, the mass of electron, the mass of Z°, and $\alpha_e$.   The calculated masses are in good agreement with the observed values.  The cosmic vacuum and the big bang universe are connected by the cosmic wormhole network everywhere all the time.  The cosmic wormhole network becomes active, when the spacetime dimensions of the cosmic vacuum and the big bang universe are compatible.  The active cosmic wormhole network allows non-zero vacuum energy from the cosmic vacuum.   The results are the inflation during the big bang, the late-time cosmic accelerating expansion, the cosmic accelerating contraction, and the final cosmic accelerating contraction.

## 1. Introduction

Supersymmetry involves a symmetrical change in spacetime labels in terms of spacetime coordinates, while internal symmetry involves a symmetrical change in non-spacetime labels in terms of charged fields, isotopic spin, and colors. In this paper, the cosmic vacuum is the wavefunction of the eleven dimensional Planck supermembrane [1]. The Planck wavefunction is the superposition of dimensions from eleven to four dimensional spacetime with decreasing energy and increasing size. The cosmic vacuum is a gigantic cosmic particle-wave. It undergoes a gigantic slow cosmic oscillation between the high-energy eleven dimensional spacetime and the low-energy four dimensional spacetime.

The origin of the ordinary (baryonic) matter in the big bang universe is the collapsed Planck wavefunction due to the lepton-quark entangled state with two unequal sets of spacetime as the quantum system-measurement entangled state in Hilbert space [2]. The collapsed Planck wavefunction has four dimensional spacetime and seven dimensional internal space (non-spacetime) with seven gauge bosons. The remnants of the cosmic oscillation are cosmic radiation, gravity, and microscopic particle-wave in the four dimensional spacetime. Internal symmetries for gauge bosons, leptons, quarks, and black holes are in the seven dimensional internal space. The result is the big bang universe with the cosmic vacuum.

The cosmic vacuum and the big bang universe are connected by the cosmic wormhole network everywhere all the time. The cosmic wormhole network becomes active, when the spacetime dimensions of the cosmic vacuum and the big bang universe are compatible. The active cosmic wormhole network allows non-zero vacuum energy from the cosmic vacuum.

In the Section 2, the unobservable cosmic vacuum involving the Planck wavefunction will be discussed. The Section 3 covers the collapse of the Planck wavefunction. The Section 4 reviews the observable big bang universe. In the Section 5, the four periods of the active cosmic wormhole network will be discussed.

## 2. The Cosmic Vacuum

A supermembrane can be described as two dimensional object that moves in an eleven dimensional space-time [1]. This supermembrane can be converted into the ten dimensional superstring with the extra dimension curled into a circle to become a closed superstring. The cosmic vacuum state is the wavefunction of this eleven-dimensional Planck supermembrane. The Planck wavefunction is the superposition of dimensions from eleven to four dimensional spacetime with decreasing energy and increasing size. It is a gigantic cosmic particle-wave. In the Planck wavefunction, each space dimension can be described by a fermion and a boson as the following hierarchy:

$$F_4\ B_4\ F_5\ B_5\ F_6\ B_6\ F_7\ B_7\ F_8\ B_8\ F_9\ B_9\ F_{10}\ B_{10}\ F_{11}\ B_{11}$$

where B and F are boson and fermion in each spacetime dimension. The probability to transforming a fermion into its boson partner in the adjacent dimension is same as the fine structure constant, $\alpha$, the probability of a fermion emitting or absorbing a boson. The probability to transforming a boson into its fermion partner in the same dimension is also the fine structure constant, $\alpha$. This hierarchy can be expressed in term of the dimension number, D,

$$M_{D-1,\ B} = M_{D,\ F}\ \alpha_{D,\ F}, \qquad (1)$$

$$M_{D,\ F} = M_{D,\ B}\ \alpha_{D,\ B}, \qquad (2)$$

where $M_{D,\ B}$ and $M_{D,F}$ are the masses for a boson and a fermion, respectively, and $\alpha_{D,\ B}$ or $\alpha_{D,F}$ is the fine structure constant, which is the ratio between the energies of a boson and its fermionic partner. All fermions and bosons are related by the order $1/\alpha$.

The cosmic vacuum is a dynamic vacuum. The cosmic vacuum undergoes a gigantic slow cosmic quantum oscillation between the high-energy small eleven dimensional spacetime and the low-energy large four dimensional spacetime. The cosmic oscillation is through decay into concerted particles that then are recombined. During the cosmic oscillation, the expansion depends only on the decay mode, and the contraction depends only on the combination mode that combine the concerted particles together. There are no force field and cosmic radiation in the cosmic vacuum.

As in a microscopic quantum system before the measurement process, there are no independent permanent components within the cosmic vacuum during the cosmic quantum oscillation. The average dimension of the particles in the cosmic vacuum changes continuously and concertedly. All particles are in a concerted change.

The decay mode in the cosmic vacuum is different from the decay mode in the big bang universe. In the decay mode in the cosmic vacuum, the decay particles are in a concerted state as in the microscopic quantum system before the measurement. All decay particles "remember" their parent Planck supermembranes, as the in the microscopic system before the measurement, all individual spin states remember their original composite spin state regardless of time and space. Instead of explosive exponential radiation decay in the big bang universe, the decay in the cosmic vacuum is extremely orderly, one at the time per parent Planck supermembrane. As soon as one particle decay, all other decay particles adjust their masses, so average mass of the particles in the cosmic vacuum varies with time. All of them are under the concerted change.

The energy as the cosmological constant in the vacuum state is (mass)$^4$ where the mass is the mass for the symmetry breaking. The vacuum energy for the cosmic vacuum is, therefore, equal to $M_p^4$ where $M_p$ is the Planck mass. Assuming a logarithm relation, the corresponding Planck time in the cosmic vacuum is equal to $T_p^{1/4}$ where $T_p$ is the Planck time ($5.39 \times 10^{-44}$ second). The Planck time in the cosmic vacuum is, therefore, equal to $1.52 \times 10^{-11}$ second. Since decay rate is related to the Planck time, the decay rate in the cosmic vacuum is one decay particle per $1.52 \times 10^{-11}$ second per parent Planck supermembrane. With this decay rate, the time span of the cosmic vacuum in one oscillation is very long. For example, it takes 27.6 billion years to decay from the eleven dimensional Planck supermembrane to the four dimensional bosons.

### *3. The Collapse of the Planck wavefunction*

The Planck wavefunction, the cosmic particle-wave, is collapsed by the entangled state with two unequal sets of spacetime as the quantum-measurement entangled state in Hilbert space [2]. It involves the crossing of two unequal sets of spacetime. One of the possible entangled states is the lepton-quark entangled state that consists of seven dimensions mostly for leptons and seven auxiliary dimensions mostly for quarks as shown in Fig. 1.

The entangled state undergoes a "quantum leap" through a cosmic wormhole network to become the collapsed Planck wavefunction as the quantum leap for the quantum system-measurement in Hilbert space. One of the collapsed Planck wavefunctions is for the ordinary (baryonic) matter in the big bang universe. The collapsed Planck wavefunction for the ordinary (baryonic) matter in the big bang universe has four dimensional spacetime and seven dimensional internal space (non-spacetime) with seven gauge bosons. Other collapsed Planck wavefunctions are the exotic cold dark matter in the big bang universe. The current density of the baryonic matter is 0.012 - 0.041, while the density of the exotic dark matter is 0.1 - 0.5 [3]. Since the baryonic matter is one of the eight possible collapsed wavefunctions, the ratio of the baryonic matter to the exotic dark matter agrees with the observed value.

After the collapse, internal symmetries are in the seven dimensional internal space, and the remnants of the cosmic oscillation are in the four dimensional spacetime. The remnants are cosmic radiation and gravity, corresponding to the expansion and the contraction in the cosmic oscillation. Cosmic radiation maximizes the distance between any two objects as in the expansion, while gravity minimizes the distance between any two objects as in the contraction. The interactions of cosmic radiation and gravity toward all materials are indiscriminate. Everything can be involved in the same expansion and the contraction. Both cosmic radiation and gravity follow the property of spacetime. Ubiquitous cosmic radiation represents spacetime, and gravity is the curvature of four dimensional spacetime. Cosmic radiation and gravity do not need internal space.

The remnant of the Planck wavefunction is also shown in microscopic particle-wave in quantum mechanics. In the wavefuncition, the spread out momentum corresponds to the continuously changing momentum in the Planck wavefunction, and the spread out location corresponds to the continuously changing size of the Planck wavefunction.   The cosmic uncertainty principle, therefore, is the uncertainty to identify a particle with precise position and momentum in the Planck wavefunction.  Microscopic particle-wave is the remnant of ever changing spacetime in the Planck wavefunction.  Microscopic particle-wave does not need internal space.   While the time scale of the Planck wavefunction is the large one-fourth power of the Planck time, the time scale of microscopic particle-wave is the small Planck time.

## *3. The Big Bang Universe*

The collapsed Planck wavefunction in the big bang universe results in four dimensional spacetime and seven dimensional internal space.  The seven dimensional internal space is the base for the gauge bosons and the periodic table of elementary particles as described in details in Reference 4.  It is briefly reviewed here.

The permanent lepton-quark entangled state requires the violations of symmetries (CP and P) in spacetime and the internal symmetry acting on the internal space.   CP nonconservation is required to distinguish the lepton-quark entangled state from the CP symmetrical cosmic radiation that is absence of the lepton-quark entangled state.  P nonconservation is required to achieve chiral symmetry for massless leptons (neutrinos), so some of the dimensional fermions can become leptons (neutrinos).  Various internal symmetry groups are require to organize leptons, quarks, black hole particles, and force fields.  The force fields include the long-range massless force to bind leptons and quarks, the short-range force to bind quarks, the short-range interactions for the flavor change among quarks and leptons, and the short-range interaction for the eleven dimensional particles inside black hole.

The lepton-quark entangled state consists of two sets of seven internal dimensional space.  The internal space mostly for leptons becomes the dominating internal space, while the internal space mostly for quarks is hidden.  Finally, the gauge bosons, leptons, and quarks absorb the common scalar field of four dimensional spacetime to acquire the common label of four dimensional spacetime through Higgs mechanism.   The results are the gauge bosons (Table 1) and the periodic table of elementary particles (Table 2).

For the big bang universe, the seven internal space dimensions are arranged in the same way as the spacetime dimensions in the cosmic vacuum.

$$F_5 \ B_5 \ F_6 \ B_6 \ F_7 \ B_7 \ F_8 \ B_8 \ F_9 \ B_9 \ F_{10} \ B_{10} \ F_{11} \ B_{11}$$

where B and F are boson and fermion in each spacetime dimension. The gauge bosons in the big bang universe can be derived from Eqs. (1) and (2).   Assuming $\alpha_{D,B} = \alpha_{D,F}$, the relation between the bosons in the adjacent dimensions, then, can be expressed in term of the dimension number, D,

$$M_{D-1, B} = M_{D, B} \; \alpha^2_D, \qquad (3)$$

where D= 6 to 11, and $E_{5,B}$ and $E_{11,B}$ are the energies for the dimension five and the dimension eleven, respectively.

The lowest energy is the Coulombic field, $E_{5,B}$

$$\begin{aligned} E_{5, B} &= \alpha \, M_{6,F} \\ &= \alpha \, M_e, \end{aligned} \qquad (4)$$

where $M_e$ is the rest energy of electron, and $\alpha = \alpha_e$, the fine structure constant for the magnetic field. The bosons generated are called "dimensional bosons" or "$B_D$". Using only $\alpha_e$, the mass of electron, the mass of $Z^0$, and the number of extra dimensions (seven), the masses of $B_D$ as the gauge boson can be calculated as shown in Table 1.

**Table 1.** The Energies of the Dimensional Bosons
$B_D$ = dimensional boson, $\alpha = \alpha_e$

| $B_D$ | $M_D$ | GeV | Gauge Boson | Interaction |
|---|---|---|---|---|
| $B_5$ | $M_e \, \alpha$ | $3.7 \times 10^{-6}$ | A | electromagnetic, U(1) |
| $B_6$ | $M_e / \alpha$ | $7 \times 10^{-2}$ | $\pi_{1/2}$ | strong, SU(3) |
| $B_7$ | $M_6 / \alpha_w^2 \cos \theta_w$ | 91.177 | $Z_L^0$ | weak (left), SU(2)$_L$ |
| $B_8$ | $M_7 / \alpha^2$ | $1.7 \times 10^6$ | $X_R$ | CP (right) nonconservation, U(1)$_R$ |
| $B_9$ | $M_8 / \alpha^2$ | $3.2 \times 10^{10}$ | $X_L$ | CP (left) nonconservation, U(1)$_L$ |
| $B_{10}$ | $M_9 / \alpha^2$ | $6.0 \times 10^{14}$ | $Z_R^0$ | weak (right), SU(2)$_R$ |
| $B_{11}$ | $M_{10} / \alpha^2$ | $1.1 \times 10^{19}$ | $G_b$ | black hole, large N color field |

In Table 1, $\alpha_w$ is not same as $\alpha$ of the rest, because there is symmetry group mixing between $B_5$ and $B_7$ as the symmetry mixing in the standard theory of the electroweak interaction, and $\sin\theta_w$ is not equal to 1. As shown in Reference 4, $B_5$, $B_6$, $B_7$, $B_8$, $B_9$, and $B_{10}$ are A (massless photon), $\pi_{1/2}$, $Z_L^0$, $X_R$, $X_L$, and $Z_R^0$, respectively, responsible for the electromagnetic field, the strong interaction, the weak (left handed) interaction, the CP (right handed) nonconservation, the CP (left handed) nonconservation, and the P (right handed) nonconservation, respectively. The calculated value for $\theta_w$ is $29.69^0$ in good agreement with $28.7^0$ for the observed value of $\theta_w$ [5].

The calculated energy for $B_{11}$ ($G_b$, black hole gluon, equivalent to the Planck supermembrane) is $1.1 \times 10^{19}$ GeV in good agreement with the Planck mass, $1.2 \times 10^{19}$ GeV. $G_b$ is the eleven dimensional gauge boson for the interaction inside the event horizon of the black hole. The confinement of all particles inside black hole is similar to the confinement of all quarks inside hardon, so the internal symmetry is based on large N color [6]. Black hole gluon corresponds to gluon, and black hole gluino, the supersymmetrical partner, corresponds to quark. Therefore, inside the horizon of black hole, the gravitational force goes zero, and the large N color field appears. The eleven dimensional black hole gluon and gluino are surrounded by the four dimensional gravity.

These eleven dimensional black hole gluons and gluinos are still in internal space (color) whose internal symmetry generates large N color force field. They are not eleven dimensional Planck supermembranes that hold together by the tangle of D-branes and strings. The event horizon is the dividing line between the four dimensional spacetime labeled force and the internal space labeled force. As described later, all black hole universe is the end of the cosmic cycle, and all Planck supermembrane is the beginning of the cosmic cycle.

The calculated masses of all gauge bosons are also in good agreement with the observed values. Most importantly, the calculation shows that exactly seven extra dimensions are needed for all fundamental interactions.

The model for leptons and quarks is shown in Fig. 1. The periodic table for elementary particles is shown in Table 2.

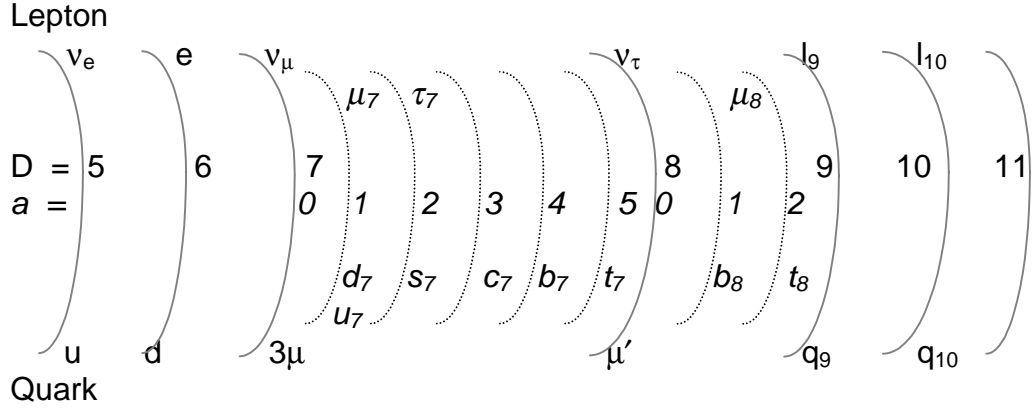

**Fig. 1.** Leptons and quarks in the dimensional orbits
D = dimensional number, a = auxiliary dimensional number

**Table 2.** The Periodic Table of Elementary Particles
D = dimensional number, a = auxiliary dimensional number

| D | A = 0 | 1 | 2 | a = 0 | 1 | 2 | 3 | 4 | 5 | |
|---|---|---|---|---|---|---|---|---|---|---|
| | Lepton | | | Quark | | | | | | Boson |
| 5 | $l_5 = \nu_e$ | | | $q_5 = u = 3\nu_e$ | | | | | | $B_5 = A$ |
| 6 | $l_6 = e$ | | | $q_6 = d = 3e$ | | | | | | $B_6 = \pi_{1/2}$ |
| 7 | $l_7 = \nu_\mu$ | $\mu_7$ | $\tau_7$ | $q_7 = 3\mu$ | $u_7/d_7$  $s_7$ | $c_7$ | $b_7$ | $t_7$ | | $B_7 = Z_L^0$ |
| 8 | $l_8 = \nu_\tau$ | $\mu_8$ | | $q_8 = \mu'$ | $b_8$  $t_8$ | | | | | $B_8 = X_R$ |
| 9 | $l_9$ | | | $q_9$ | | | | | | $B_9 = X_L$ |
| 10 | $F_{10}$ | | | | | | | | | $B_{10} = Z_R^0$ |
| 11 | $F_{11}$ | | | | | | | | | $B_{11} = G_b$ |

D is the dimensional orbital number for the seven extra space dimensions. The auxiliary dimensional orbital number, a, is for the seven extra auxiliary space dimensions, mostly for quarks. All gauge bosons, leptons, and quarks are located on the seven dimensional orbits and seven auxiliary orbits. Most leptons are dimensional fermions, while all quarks are the sums of subquarks.

The fermion mass formula for massive leptons and quarks is derived from Reference 4 as follows.

$$M_{F_{D,a}} = \sum M_{F_{D,0}} + M_{AF_{D,a}}$$

$$= \sum M_{F_{D,0}} + \frac{3}{2} M_{B_{D-1,0}} \sum_{a=0}^{a} a^4 \quad (5)$$

$$= \sum M_{F_{D,0}} + \frac{3}{2} M_{F_{D,0}} \alpha_D \sum_{a=0}^{a} a^4$$

Each fermion can be defined by dimensional numbers (D's) and auxiliary dimensional numbers (a's). The compositions and calculated masses of leptons and quarks are listed in Table 3.

**Table 3.** The Compositions and the Constituent Masses of Leptons and Quarks
D = dimensional number and a = auxiliary dimensional number

|         | $D_a$                        | Composition                              | Calc. Mass       |
|---------|------------------------------|------------------------------------------|------------------|
| Leptons | $D_a$ for leptons            |                                          |                  |
| $\nu_e$ | $5_0$                        | $\nu_e$                                  | 0                |
| e       | $6_0$                        | e                                        | 0.51 MeV (given) |
| $\nu_\mu$ | $7_0$                      | $\nu_\mu$                                | 0                |
| $\nu_\tau$ | $8_0$                     | $\nu_\tau$                               | 0                |
| $\mu$   | $6_0 + 7_0 + 7_1$            | $e + \nu_\mu + \mu_7$                    | 105.6 MeV        |
| $\tau$  | $6_0 + 7_0 + 7_2$            | $e + \nu_\mu + \tau_7$                   | 1786 MeV         |
| $\mu'$  | $6_0 + 7_0 + 7_2 + 8_0 + 8_1$ | $e + \nu_\mu + \mu_7 + \nu_\tau + \mu_8$ | 136.9 GeV        |
| Quarks  | $D_a$ for quarks             |                                          |                  |
| u       | $5_0 + 7_0 + 7_1$            | $u_5 + q_7 + u_7$                        | 330.8 MeV        |
| d       | $6_0 + 7_0 + 7_1$            | $d_6 + q_7 + d_7$                        | 332.3 MeV        |
| s       | $6_0 + 7_0 + 7_2$            | $d_6 + q_7 + s_7$                        | 558 MeV          |
| c       | $5_0 + 7_0 + 7_3$            | $u_5 + q_7 + c_7$                        | 1701 MeV         |
| b       | $6_0 + 7_0 + 7_4$            | $d_6 + q_7 + b_7$                        | 5318 MeV         |
| t       | $5_0 + 7_0 + 7_5 + 8_0 + 8_2$ | $u_5 + q_7 + t_7 + q_8 + t_8$           | 176.5 GeV        |

The calculated masses are in good agreement with the observed constituent masses of leptons and quarks [7,8]. The mass of the top quark found by Collider Detector Facility is 176 ± 13 GeV [7] in a good agreement with the calculated value, 176.5 GeV. The masses of elementary particles can be calculated with only four known constants: the number of the extra spatial dimensions in the supermembrane, the mass of electron, the mass of Z°, and $\alpha_e$. The calculated masses are in good agreement with the observed values.

## 5. *The Cosmic Wormhole network*

The cosmic vacuum and the big bang universe are connected by the cosmic wormhole network everywhere all the time. When the spacetime dimensions in the two universes are different, the cosmic wormhole network remains dormant, and the vacuum energy (cosmological constant) with respect to the big bang universe is zero. When the spacetime dimensions in the two universes are compatible, the cosmic wormhole network becomes active. During this period, the active wormhole network allows non-zero vacuum energy. The positive vacuum energy causes negative pressure, inducing anti-gravity input, while the negative vacuum energy causes positive pressure, inducing gravity input.

During one cycle of the universe, there are four periods of active cosmic wormhole network.  The first period of active cosmic wormhole network is the inflation induced by anti-gravity input at the start of the expansion of the universe.  The Planck supermembranes that decay in a short time into internal symmetry four-plus-seven dimensional universe.  As soon as the decay takes place, the cosmic wormhole network becomes dormant due to the difference in the spacetime dimensions between the two universes.

The dormant cosmic wormhole network becomes active when the energy level of the decay mode in the cosmic vacuum is equal to the energy level between the five-dimensional spacetime and the four dimensional spacetime as in the big bang universe.  The vacuum energy in the big bang universe again starts to become positive, inducing the negative pressure and anti-gravity input, causing accelerating expansion.  It is the second period of active cosmic wormhole network, which is happening at the present as late-time cosmic accelerating expansion.

The evidence for late-time cosmic accelerating expansion is from the recent observations of large-scale structure that suggests that the universe is undergoing cosmic accelerating expansion, and it is assumed that the universe is dominated by a dark energy with negative pressure recently [9].  The dark energy can be provided by a non-vanishing cosmological constant or quintessence [10], a scalar field with negative pressure.  However, a cosmological constant requires extremely fine-tuned [11].  Quintessence requires an explanation for the late- time cosmic accelerating expansion [12].  Why does quintessence dominate the universe only recently?   One of the explanations is the *k*-essence model where the pressure of quintessence switched to a negative value at the onset of matter-domination in the universe [12].

According to the dynamic cosmic vacuum model, this late-time cosmic accelerating expansion is caused by the anti-gravity input during the second period of active cosmic wormhole network.  The compatible energy level for the cosmic vacuum and the big bang universe is between the energy levels of the five dimensional fermion and the four dimensional boson.  From Eqs (1) and (2) and Table 1, the energies of the five dimensional fermion and the four dimensional boson are calculated to be $2.72 \times 10^{-8}$ GeV and $1.99 \times 10^{-10}$ GeV, respectively.  The energy of the Planck supermembrane is $1.1 \times 10^{19}$ GeV from Table 1.  The energy ratio between the Planck supermembrane and the five dimensional fermion and the four dimensional boson are $4.17 \times 10^{26}$ and $5.71 \times 10^{28}$, respectively.   In other words, there are $4.17 \times 10^{26}$ five dimensional fermions and $5.71 \times 10^{28}$ four dimensional bosons per Planck supermembrane.  The decay rate is one particle per one-fourth power of the Planck time ($1.52 \times 10^{-11}$ second) per parent Planck supermembrane.  The total time for the decay is 0.2 billion years for the five dimensional fermion, and 27.6 billion years for the four dimensional boson.  Therefore, anti-gravity input starts in 0.2 billion years after the Big Bang, and ends in 27.6 billions years, and anti-gravity input causes accelerating expansion in the big bang universe.

If the relative rate of the rise and the fall of anti-gravity input in the big bang universe during this period follows a smooth curve with respect to time, the maximum relative rate (100) of anti-gravity input occurs in about 13.9 billion years after the big bang. It is coincidentally about the age of the universe. The relative rate of anti-gravity input versus time is shown in Fig. 2.

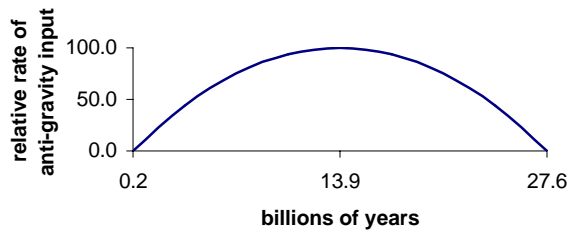

**Fig. 2.** relative rate of anti-gravity input in the big bang universe versus time

In other words, the greatest acceleration occurs at the peak of the space absorption in 13.9 billion years after the big bang.

The oscillation of the cosmic vacuum now is in the expansion mode. The cosmic vacuum may expand all the way down to the beginning of the two dimensional spacetime, or may expand only down to the beginning of the four dimensional spacetime, that is about 14 billion years from now. When the oscillation is in the contraction mode in the future, and the cosmic vacuum again passes the four dimensional spacetime, the big bang universe will gain gravity input to pay back anti-gravity input during the second period of the active cosmic wormhole network. The result is the accelerating contraction of the big bang universe. This third period of the active cosmic wormhole network prevents the big bang universe to expand forever with an infinite large space. Therefore, the big bang universe and the cosmic vacuum can contract together.

Toward the end of the contraction mode, black holes become the dominating feature in the big bang universe. Black hole contains eleven dimensional black hole gluon and gluino surrounded by gravity. When both universes are near eleven dimensional spacetime, they again become compatible. In this last period of active cosmic wormhole network, the big bang universe again gains gravity input. The result is to gather all material in the big bang universe for the final contraction into the eleven dimensional black hole gluons and gluinos surrounded by four dimensional gravity in one area.

Moment after the final accelerating contraction, anti-gravity input starts to start the inflation. The anti-gravity input annihilates gravity surrounding the black hole gluons and gluinos. Without gravity, black hole gluons and gluinos disappear as quarks disappear outside of hadrons. What remain are the eleven dimensional Planck supermembranes that hold together by the tangle of D-branes and strings.  Immediately, the Planck wavefunction of the eleven dimensional Planck supermembranes collapses, and another cosmic cycle starts.  The inflation, therefore, is not an add-on but a necessity to start the cosmic cycle.

The particles derived from the very first collapsed Planck wavefunction came from the same quantum system, the cosmic vacuum, so they were in a completely equilibrium state.  The universe in the first cosmic cycle was smooth with few distinctive structures.  After few cosmic cycles, the universe develops wrinkles.  The current big bang universe has many distinctive structures, indicating the history of few cosmic cycles.

## *6.    Conclusion*

Supersymmetry involves a symmetrical change in spacetime labels in terms of spacetime coordinates, while internal symmetry involves a symmetrical change in non-spacetime labels in terms of charged fields, isotopic spin, and colors. In this paper, the cosmic vacuum is the wavefunction of the eleven dimensional Planck supermembrane .  The Planck wavefunction is the superposition of dimensions from eleven to four dimensional spacetime with decreasing energy and increasing size.  The cosmic vacuum is a gigantic cosmic particle-wave.  It undergoes a gigantic slow cosmic oscillation between the high-energy eleven dimensional spacetime and the low-energy four dimensional spacetime.

The origin of the big bang universe is the lepton-quark entangled state with two unequal sets of spacetime as the quantum system-measurement entangled state in Hilbert space.  The entangled state undergoes a "quantum leap" through a cosmic wormhole network to become the collapsed Planck wavefunction.  The collapsed Planck wavefunction has four dimensional spacetime and seven dimensional internal space (non-spacetime) with seven gauge bosons. The remnants of the cosmic oscillation are cosmic radiation, gravity, and microscopic particle-wave in the four dimensional spacetime. Internal symmetries for gauge bosons, leptons, quarks, and black holes are in the seven dimensional internal space.  The result is the big bang universe with the cosmic vacuum.  There is no curled-up space dimension and compacitification. Except at the very beginning of the cosmic cycle and black hole, the big bang universe has pure four-dimensional spacetime. The cosmic vacuum has the four-to-eleven dimensional spacetime supersymmetry, while the big bang universe has the four dimensional spacetime and the seven dimensional internal space.

The gauge bosons and the periodic table of elementary particles can be derived from the collapsed Planck wavefunction. There is black hole gluon specifically for the interaction inside black hole. The masses of gauge bosons and elementary particles can be calculated  with only four known constants: the number of the extra spatial dimensions in the supermembrane, the mass of electron, the mass of Z°, and $\alpha_e$. The calculated masses are in good agreement with the observed values.

The cosmic vacuum and the big bang universe are connected by the cosmic wormhole network everywhere all the time. The cosmic wormhole network becomes active, when the spacetime dimensions of the cosmic vacuum and the big bang universe are compatible. The active cosmic wormhole network allows non-zero vacuum energy from the cosmic vacuum. The results are the inflation during the big bang, the late-time cosmic accelerating expansion, the cosmic accelerating contraction, and the final cosmic accelerating contraction.

The big bang universe cycle starts with the collapsed Planck wavefunction of the Planck supermembrane, and ends with all black hole universe for the big bang universe. The inflation is not an add-on but a necessity to start the cosmic cycle. According to the calculation, we are near the peak of the late-time cosmic accelerating expansion.